# Development of an Acoustic Transceiver for Positioning Systems in Underwater Neutrino Telescopes

Giuseppina Larosa[a], Miguel Ardid[a], Carlos D. Llorens[b], Manuel Bou-Cabo[a],
Juan A. Martínez-Mora[a], Silvia Adrián-Martínez[a]

[a] Institut d'Investigació per a la Gestió Integrada de les Zones Costaneres (IGIC) – Universitat Politècnica de València,
C/ Paranimf 1, 46730 Gandia, València, SPAIN
e-mail: giula@doctor.upv.es

[b] E.P.S. Gandia, Universitat Politècnica de València, C/ Paranimf 1, 46730 Gandia, València, SPAIN

*Abstract*— **In this paper, we present the acoustic transceiver developed for the positioning system in underwater neutrino telescopes. These infrastructures are not completely rigid and need a positioning system in order to monitor the position of the optical sensors of the telescope which have some degree of motion due to sea currents. To have a highly reliable and versatile system in the infrastructure, the transceiver has the requirements of reduced cost, low power consumption, high intensity for emission, low intrinsic noise, arbitrary signals for emission and the capacity of acquiring and processing the received signal on the board. The solution proposed and presented here consists of an acoustic transducer that works in the 20-40 kHz region and withstands high pressures (up to 500 bars). The electronic-board can be configured from shore and is able to feed the transducer with arbitrary signals and to control the transmitted and received signals with very good timing precision. The results of the different tests done on the transceiver in the laboratory are described here, as well as the change implemented for its integration in the Instrumentation Line of ANTARES for the in situ tests. We consider the transceiver design is so versatile that it may be used in other kinds of marine positioning systems, alone or combined with other marine systems, or integrated in different Earth-Sea Observatories, where the localization of the sensors is an issue.**

## I. Introduction

The acoustic transceiver presented in the article has been developed to be used in the acoustic positioning system of the neutrino telescopes in the Mediterranean Sea KM3NeT [1], and it is going to be tested on the ANTARES neutrino telescope. ANTARES is currently the biggest underwater neutrino telescope in the world and in operation in the Northern Hemisphere [2−3]. The detector is located in the Mediterranean Sea on a marine site 40 km SE offshore from the city of Toulon (France), at about 2400 m depth. Its construction was completed in May 2008, and now it is collecting data, which is analyzed in order to bring insights into different scientific problems related not only to astroparticles, but also in several fields of Earth-Sea Sciences. On the other hand, KM3NeT is a European Consortium that aims to design, build and operate a cubic kilometre neutrino telescope in the Mediterranean Sea [1−4]. The project is now in the Preparatory Phase for the Construction funded by the VII Framework Program of the European Union.

Undersea neutrino telescopes have become very important tools for the study of the Universe. Moreover, these infrastructures are also abyssal multidisciplinary observatories with the installation of specialized instrumentation for biology, seismology, gravimetry, radioactivity, geomagnetism, oceanography and geochemistry offering a unique opportunity to explore the properties of a deep Mediterranean Sea site over a period of many years. The main elements of a neutrino telescope are an array of optical sensors (photomultipliers in glass spheres) located in flexible structures deployed in the deep sea and maintained vertical with buoys. For KM3NeT, the array will cover large volumes, of the order of a cubic kilometre, to have adequate sensitivity for the expected fluxes of neutrinos in these processes. It is able to detect the Cherenkov light from the muons produced by neutrino interactions with matter around the detector. The arrival times of the light collected by the optical detectors can be used to reconstruct the muon trajectory, and consequently that of the neutrino. The accuracy of the reconstruction of the muon track depends on the precision in measurement of light arrival time and on precise knowledge of the positions of the optical detectors [5−7]. The positioning system is necessary because marine currents may produce inclination of the structures, and thus displacement the optical sensors of the telescope by several metres from the nominal position. Then, the precise knowledge of the relative positions of all optical sensors is essential for a good operation of the telescope, and must be known with ~10 cm accuracy. On the other hand, the absolute geo-referenced positions are needed to point back to astronomical sources. In order to know and monitor with precision the relative positions of the optical modules an triangulation method is applied in the acoustic positioning system constituted of receiving hydrophones attached to the structures and of emitting transceivers in fixed positions near the sea bottom.

A considerable effort has been made by the KM3NeT Collaboration for the development of such system [8−10]. Here, we will present the work done in order to develop, test and integrate the solution for the KM3NeT acoustic transceiver.

The transceiver design is very versatile, and thus, it can be easily adapted to other kinds of marine positioning systems, alone or combined with other marine systems, or integrated in different Earth-Sea Observatories, where the localization of the sensors is an issue. Therefore, we think that it can be a very useful tool in geo-processing applications in marine environment.

In Section II, the acoustic transceiver is described. The laboratory tests performed on it are discussed in Section III. Section IV shows the activities for the integration of the system in the ANTARES telescope for the in situ tests. Finally, the conclusions are presented, as well as the different possibilities of the system for being used in other marine positioning or localization systems.

## II. THE ACOUSTIC TRANSCEIVER

The Acoustic Positioning System (APS) for the future KM3NeT neutrino telescope consists of a series of acoustic transceivers distributed on the sea bottom and receivers located on the lines near the optical modules. Each of these acoustic transceivers is composed of a transducer and one electronic board named '*sound emission board*'. Next, we will present these two parts of our system.

### A. The acoustic sensor

The acoustic sensor has been selected to attend to the specifications needed for the KM3NeT positioning: withstand high pressure, good receiving sensitivity and transmitting power, nearly omnidirectional, low electronic noise, high reliability, and affordable for the units needed in a cubic kilometer. Among different options we have selected the Free Flooded Ring (FFR) transducers, model SX30 manufactured by Sensor Technology Ltd. The FFR transducers have geometrical forms such as rings, and then the hydrostatic pressure is the same on the inside and the outside. This characteristic form reduces the change of the properties of piezoelectric ceramic under high hydrostatic pressure. So they are a good solution to the deep submergence problem [11]. The SX30 FFRs are efficient transducers that provide reasonable power levels over wide range of frequencies and deep ocean capability. They work in the 20–40 kHz frequency range and have dimensions of 4.4 cm outer diameter, 2 cm inner diameter, 2.5 cm height.

They have unlimited depth for operation (already tested up to 440 bars [12]) with a transmitting and receiving voltage response of 133 dB Ref. 1µPa/V at 1m and -193 dB Ref. 1V/µPa, respectively. The maximum input power is 300 W with 2% duty cycle. These transducers are simple radiators and have omnidirectional directivity pattern in the plan perpendicular to the axis of the ring (plane XY), while the directivity in the other planes depends on the length of the cylinder (plane XZ), 60° for the SX30 model [13]. The cable on the free-flooded rings is 20 AWG, TPE (Thermoplastic elastomer) insulated. The cable is affixed directly to the ceramic crystal. The whole assembly is then directly coated with epoxy resin. Both the epoxy resin and the cable are stable in salt water, oils, mild acids and bases.

The cables are therefore not water blocked (fluid penetration into the cable may cause irreversible damage to the transducer). For this reason the FFR hydrophones have been over-molded with polyurethane material to block water and to facilitate its fixing and integration on mechanical structures. Figure 1 shows pictures of the FFR transducer without and with over-molding.

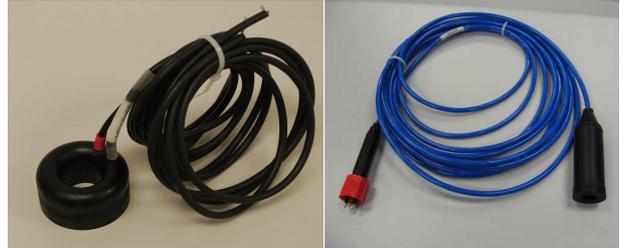

Figure 1: View of the Free Flooded Ring hydrophone (without and with over-molding).

In the next plots (Figures 2 to 5), we present the results of the tests carried out in our laboratory to characterize the transducers in terms of the transmitting and receiving voltage responses as a function of the frequency and as a function of the angle (directivity pattern). For the tests omnidirectional transducers, model ITC-1042 and calibrated RESON-TC4014 have been used as reference emitter and receiver, respectively. Particularly, Figure 2 and Figure 3 show the Transmitting Voltage Response and the Receiving Voltage Response, respectively, of the FFR hydrophones as a function of the frequency (measured in the plane XY, that is, in the perpendicular of the axis of the transducer); Figure 4 and Figure 5 show the Transmitting Voltage Response and the Receiving Voltage Response, respectively, of the FFR hydrophones as a function of the angle using a 30 kHz tone burst signal (measured in the plane XZ, 0°, which corresponds to the direction opposite to cables).

### B. The Sound Emission Board

We have developed dedicated electronics, Sound Emission Board (SEB), in order to be able to communicate, configure the transceiver and control the emission and reception. Relative to the emission, it is able to feed the signals for positioning and amplify them in order to have enough acoustic power so they could be detected from acoustic receivers at about 1 km away from the emitter.

Moreover, it stores the energy and gives enough power for the emission and to switch between emission and reception modes. The solution adopted is specially adapted to the FFR transducers and is able to feed the transducer with high amplitude short signals (a few ms) with arbitrary waveform. It has as well the capacity of acquiring the received signal. The diagram of the board prototype is shown in Figure 6. It consists of three parts: the communication and control which contains the micro-controller dsPIC (blue part), the emission part constituted by the digital amplification plus the transducer impedance matching (red part) and the reception part (green part). In the reception part a relay controlled by the dsPIC switches the mode and feeds the signal from the transducer to the receiving board of the positioning system.

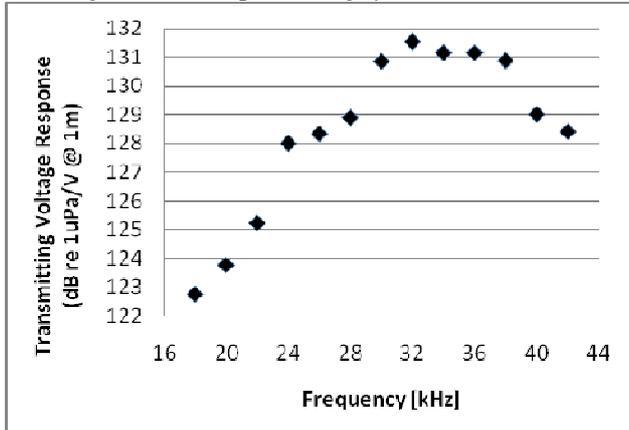
Figure 2: Transmitting Voltage Response of the FFR hydrophones.

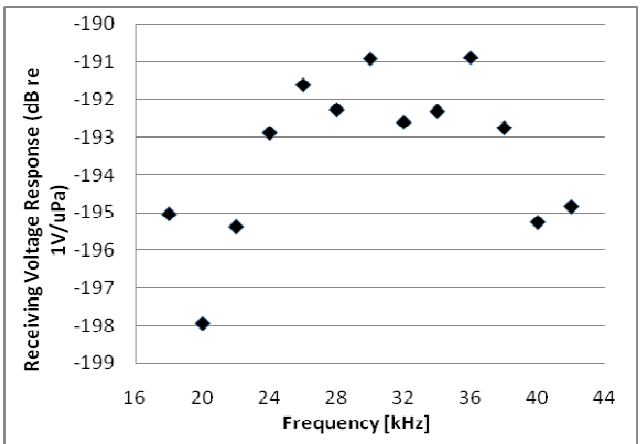
Figure 3: Receiving Voltage Response of the FFR hydrophones.

The SEB has been designed for low-power consumption and it is adapted to the neutrino infrastructure using power supplies of 12 V and 5 V with a consumption of 1 mA and 100 mA respectively, furnished by the electronic of the neutrino telescope. To avoid initial high currents, there is a current limit of 15 mA when the capacitor starts to charge, but few seconds later the current stabilizes at 1 mA. With this, a capacitor with a very low equivalent series resistance and 22mF of capacity is charged allowing storing the energy for the emission. The charge of this capacitor is monitored using the input of the ADC of the micro-controller. Moreover, the output of the micro-controller is connected through 2x Full Mosfet Driver and a MOSFET full bridge; this is successively connected at the transformer with a frequency and duty cycle programmed through the micro-controller. The transformer is able to convert the voltage of the input signal of $24V_{pp}$ to an output signal of about $500V_{pp}$.

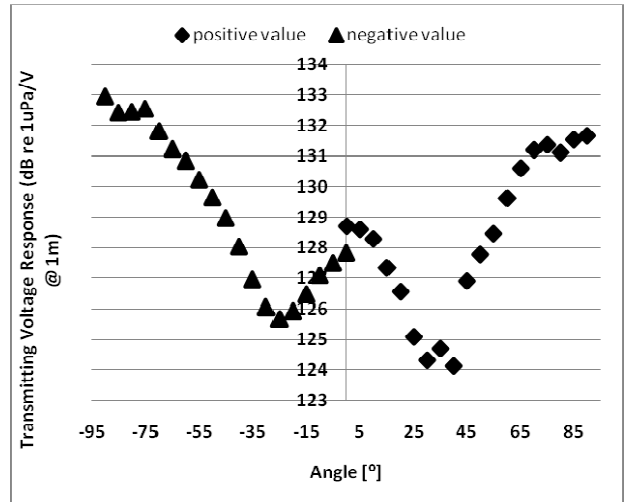
Figure 4: Transmitting Voltage Response of the FFR hydrophones.

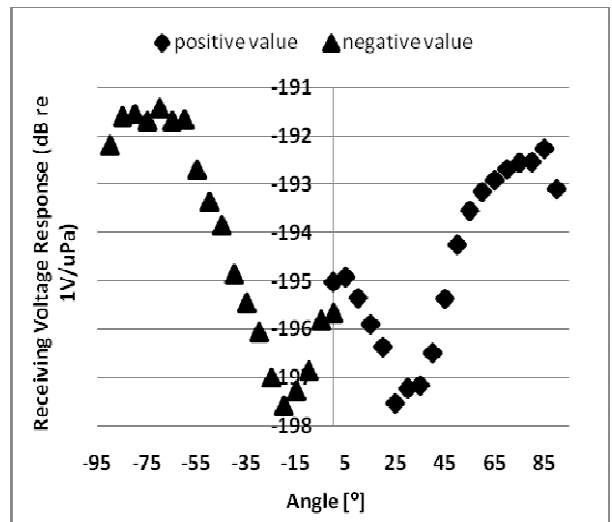
Figure 5: Receiving Voltage Response of the FFR hydrophones.

Besides, concerning the reception part of the board, the board has the possibility to directly apply an anti-aliasing filter and return the signal to an ADC of the microcontroller.

This functionality may be very interesting not only in the frame of the neutrino telescopes, but also to have the receiver implemented in different underwater applications, such as affordable sonar systems or echo-sounds.

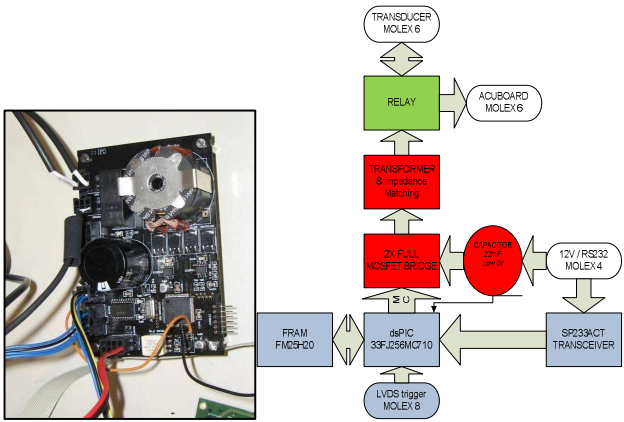

Figure 6: View and diagram of the Sound Emission Board.

The micro-controller contains the program for the emission of the signals and all the parts of control of the board. The carry frequency of the emission signal is 400 kHz and has tested up to 1-1.25MHz. The signal modulation is done with Pulse-Width Modulation technique which permits the emission of arbitrary intense short signals [14].

The basic idea of this technique is to modulate the signal digitally at a higher frequency using different width of pulses and the lower frequency signal is recovered using a low-pass filter. In addition, it will have an increase in the amplitude of the signal using a full H-Bridge. The communication of the board with the PC is established through the standard protocol RS232 using an adapter SP233 in the board. In order to have very good timing synchronization the emission is triggered using a LVDS signal.

In summary, the board, designed for an easy integration in neutrino telescope infrastructures, can be configured from shore and can emit arbitrary intense short signals or act as receiver with very good timing precision (the measured latency is 7 μs with a stability better than 1 μs), as shown in the joint tests of the INFN-CNRS-UPV acoustic positioning system for KM3NeT [15].

### III. LABORATORY TESTS OF THE TRANSCEIVER

The transceiver has been tested in the laboratory and it has been integrated in the instrumentation line of the ANTARES neutrino telescope for the in situ tests. Next, we describe briefly the activities and results of these tests.

The measurement tests in the laboratory have been performed firstly in a tank of 87.5 x 113 x 56.5 $cm^3$ with fresh-water, and secondly in a pool of 6.3 m length, 3.6 m width and 1.5 m depth. We have tested the system using the FFR hydrophone over-molded and the SEB. The molding of the transducer has been done by McArtney-EurOceanique SAS which over-molded completely the back of the transducer. Moreover, 10 meters of the cable type 4021 has been molded onto free issued hydrophones plus one connector type OM2M with its locking sleeves type DLSA-M/F. The moldings are done in polyurethane, the connector body in neoprene and the locking sleeve is in plastic.

Besides, some changes in the SEB board have been done to integrate the system in the ANTARES neutrino telescope and to test the system in situ at 2475 m depth. For simplicity and limitations in the instrumentation line, it was decided to test the transceiver only as emitter, the functionality as receiver will be tested in other in situ KM3NeT tests. The changes done in the SEB are the following: to eliminate the reception part, to adapt the RS232 connection to RS485 connection and to implement the instructions to select the kind of signals to emit matching the procedures of the ANTARES DAQ system.

To test the system we have used the transceiver in different emission configurations in combination with omnidirectional transducers, models ITC-1042 and a calibrated RESON-TC4014, used as emitter and receiver respectively. Different signals have been used (tone burst, sine sweeps, MLS signals, etc.) to see the performance of the transducer under different situations.

Figure 7 shows the Transmitting Acoustic Power of the transceiver as a function of the frequency (measured in the plane XY, that is, in the perpendicular of the axis of the transducer). The Transmitting Acoustic Power of the transceiver as a function of the angle (directivity pattern) using a 30 kHz short tone burst signal (measured in the plane XZ, 0 ° corresponds to the direction opposite to cables) is shown in Figure 8.

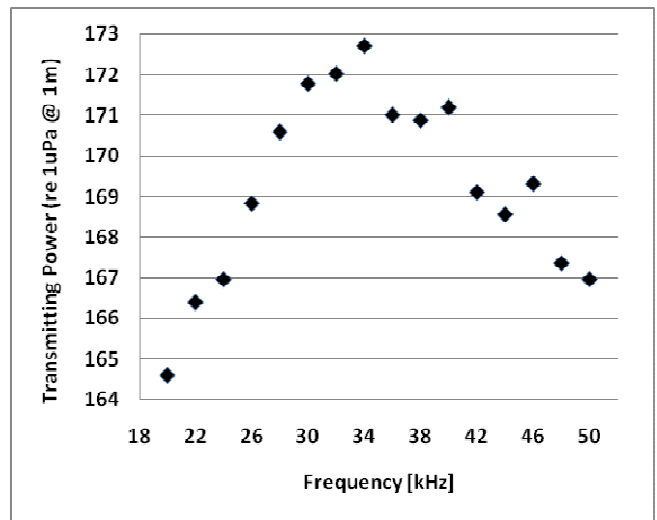

Figure 7: Transmitting Acoustic Power of the transceiver.

If we compare the Receiving and Transmitting Voltage Response of the FFR over-molded with the FFR without over-molding a loss of ~1-2 dB is observed. Figures 7 and 8 show that the results for the transmitting acoustic power in the 20-50 kHz frequency range is in the 165-173 dB re. 1μPa@1m range, in agreement with the electronics design and the specifications needed. Despite this, acoustic transmitting power may be considered low in comparison with the ones used in Long Base Line positioning systems, which usually reach values of 180 dB re. 1μPa@1m, the use of longer signals in combination with a broadband frequency range and signal processing techniques will allow us to increase the signal-to-noise ratio, and having an acoustic positioning system with about 1 μs accuracy (~ 1.5 mm) over distances of about 1 km, using less acoustic power, that is, minimizing the acoustic pollution.

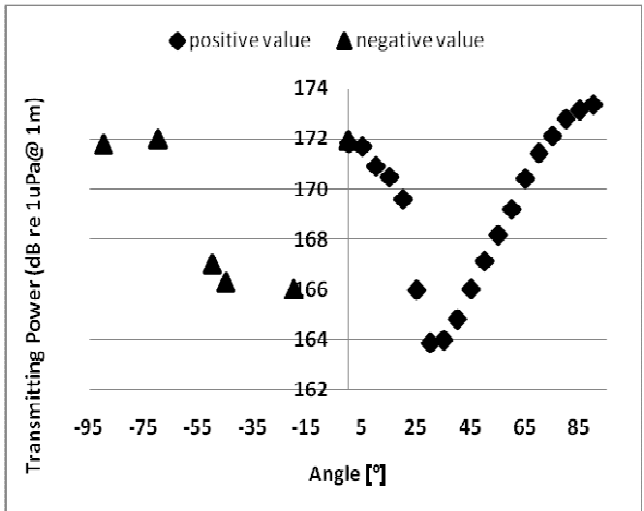

Figure 8: Transmitting Acoustic Power of the transceiver.

IV. INTEGRATION IN THE ANTARES NEUTRINO TELESCOPE

The system tested was finally integrated in the active anchor of the Instrumentation Line of ANTARES through the Laser Container used for timing calibration purposes. In fact, a new functionality for the microcontroller was implemented (to control the laser emission as well). The FFR hydrophone was fixed in the base of the line at 50 cm from the standard emitter transducer of the ANTARES positioning system with the free area of the hydrophone looking upwards. It has been fixed through a support of polyethylene designed and produced at the *Instituto de Física Corpuscular*, Valencia (Spain). The SEB was located inside a titanium container holding other electronic parts and the laser. Figures 9 and 10 show some pictures of the final integration of the system in the anchor of the Instrumentation Line of ANTARES. Finally, the Instrumentation Line was successfully deployed at 2475 m depth on 7th June 2011 at the nominal target position. The connection of the Line to the Junction Box will be in autumn 2011, when the ROV will be available, and afterwards the transceiver will be fully tested in real conditions.

V. SUMMARY AND CONCLUSIONS

We have discussed the needs of the acoustic positioning system in underwater neutrino telescopes, and presented the acoustic transceiver developed at UPV for the positioning system of KM3NeT. We have shown the results of the tests and measurements done to the FFR hydrophones and to the SEB associated, concluding that the transceiver proposed can be a good solution with the requirements and accuracy needed for such a positioning system. The transceiver, with low power consumption, is able to have a transmitting power above 170 dB ref. 1μPa@1m that combined with signal processing techniques allows to deal with the large distances involved in a neutrino telescope. Moreover, the changes performed in the transceiver, particularly in the SEB, show the capacity to adapt the electronic parts to the situation and available conditions.

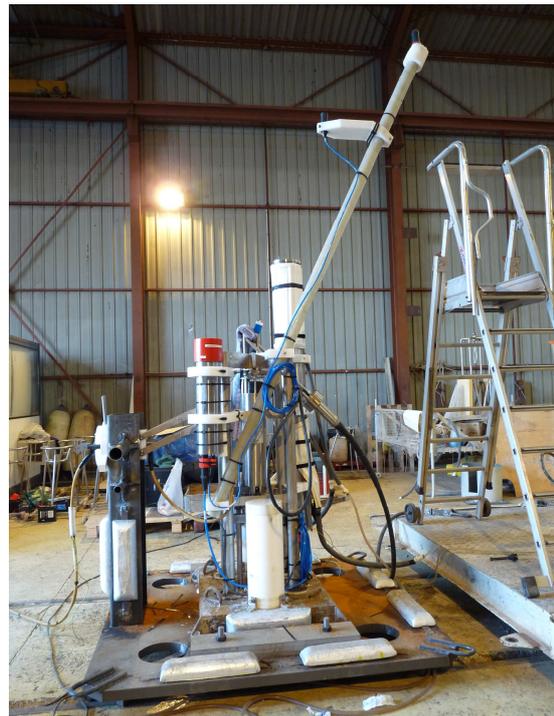

Figure 9: Picture of the anchor of the Instrumentation Line of ANTARES with the final integration of the transceiver.

The system has been integrated in the ANTARES neutrino telescope and now, we are waiting for the connection of the Instrumentation Line to test the transceiver in situ.

Finally, we would like to remark that the acoustic system proposed is compatible with the different options for the receiver hydrophones proposed for KM3NeT and it is versatile, so in addition to the positioning functionality, it can be used for acoustic detection of neutrinos studies or for bioacoustic monitoring of the sea.

Moreover, the transceiver (with slight modification) may be used in other marine positioning systems, alone or combined with other marine systems, or integrated in different Earth-Sea Observatories, where the localization of the sensors is an issue. In that sense, the experience gained from this research can be of great use for other possible applications.

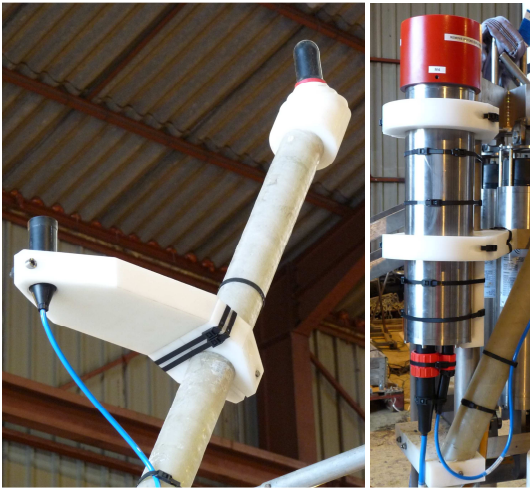

Figure 10: Views of the FFR hydrophone with the support and of the titanium laser container that contains the SEB.

ACKNOWLEDGMENTS

This work has been supported by the Ministerio de Ciencia e Innovación (Spanish Government), project references FPA2009-13983-C02-02, ACI2009-1067, AIC10-00583, Consolider-Ingenio Multidark (CSD2009-00064). It has also been funded by Generalitat Valenciana, Prometeo/2009/26, and the European 7th Framework Programme, grant no. 212525.